%% file: main.tex
\documentclass[conference,10pt]{IEEEtran}

\usepackage{cite}
\usepackage{amsmath,amssymb,amsfonts}
\usepackage{algorithmic}
\usepackage{graphicx}
\usepackage{textcomp}
\usepackage{xcolor}
\usepackage{bm}
\usepackage{comment}
\usepackage{float}
\usepackage{svg}
\usepackage{color, soul}
\usepackage{balance}
\usepackage{siunitx}
\usepackage{enumitem}
\usepackage{mathtools}
\usepackage[acronym,shortcuts]{glossaries}
\usepackage{siunitx}
\input{myacro}

\usepackage{pgfplots}
\pgfplotsset{compat=newest}
\pgfplotsset{plot coordinates/math parser=false}
\newlength\fwidth
\newlength\fheight
\pgfplotsset{ tick label style={font=\scriptsize}, label style={font=\scriptsize}, legend style={font=\scriptsize},
}
 
\title{Secret Key Generation On Aerial Rician Fading Channels Against A Curious Receiver}
\author{Mattia Piana,~\IEEEmembership{Graduate Student~Member,~IEEE}, and Stefano~Tomasin,~\IEEEmembership{Senior~Member,~IEEE}

\thanks{This work was supported by the Horizon Europe/JU SNS project ROBUST-6G (Grant Agreement no. 101139068).} \\
\small
Dept. of Information Engineering, University of Padova, Italy \\
mattia.piana@phd.unipd.it and stefano.tomasin@unipd.it}

\IEEEoverridecommandlockouts 
\begin{document}
\maketitle
\begin{abstract}
Secret key generation at the physical layer is expected to be a fundamental enabler for next-generation networks. We consider a network where the user equipment is a drone and propose a novel secret key generation solution when the eavesdropper is another node belonging to the network (curious device). We exploit drone mobility over realistic Rician fading channels. In our protocol, after a prior training phase, drone Alice chooses a trajectory of positions in space and transmits a message to Bob, on the ground, from each position. From the received messages, Bob estimates the channel gain from which a secret key is extracted. The choice of the positions is made to maximize a lower bound on the secret key capacity. Numerical simulations are used to prove the effectiveness of the proposed approach.
\end{abstract}


\section{Introduction}


\Ac{skg} at the physical layer is a mechanism that enables two devices to agree (or refresh) a stream of bits (key) that remains secret to other eavesdropping devices. Such a key can then be used to support cryptographic techniques, e.g., to achieve confidential transmissions, and provide authentication mechanisms. Two main approaches are available for \ac{skg} at the physical layer: the source-based \ac{skg}, where the randomness to generate the key is provided by the channel over which communication occurs; the channel-based \ac{skg} where instead one of the two parties transmits a random key that is kept secret from the eavesdropper as its channel does not allow to infer the key properly (e.g., the attacker channel is more noisy than the legitimate one) \cite{bloch2008physical}.
Here we focus on the channel-model technique. 

Conventional source-model \ac{skg} mechanisms at the physical layer typically exploit two characteristics of the wireless channel: reciprocity and fading \cite{aldaghri2020physical}. Usually, the former is guaranteed by both the reciprocity theory for antennas/electromagnetic propagation and the synchronization of the devices whose communication must occur within the coherence time of the channel~\cite{wilson2007channel,aldaghri2020physical}. The latter, on the other hand, is often guaranteed by random reflections of the signal in the environment. 

Yet, with poor scattering and/or slow fading, typical for example of \ac{uav} applications, \ac{skg} is challenging \cite{assaf2023high} as the \ac{los} component might be dominant \cite{lin2021secret} thus the channel can be, in the worst case, deterministic. Few recent studies tackle \ac{skg} in \ac{uav} contexts, in particular, \cite{lin2021secret} exploits \ac{mimo} and three-dimensional (3D) spatial angles to extract keys in \ac{los} environments. The authors in \cite{assaf2023high} tackle this issue by injecting randomness at the transmitter, thus creating an artificial frequency-selective fading channel. This approach is used in \cite{jangsher2023group} in a context of swarms of multiple \acp{uav} agreeing on a common key. Still in the context of single-antenna devices, \cite{lin2021secret} cannot be applied as the angle of arrival cannot be estimated, while in  \cite{assaf2023high}, provable unclonable functions (PUF) have been used to authenticate drone transmissions.

In this paper, we consider a realistic Rician fading channel for UAV transmissions (see~\cite{sharma2018study}) and propose a novel solution that exploits the mobility of the \acp{uav} to perform \ac{skg} at the physical layer, allowing a robust yet effective key agreement in low-scattering environments in the presence of a curious device. In particular, the eavesdropping device Eve belongs to the same network and has a fixed position on the ground, known to the other devices. In detail, our protocol starts with drone Alice moving in space to gather information on the environment, more precisely on the channel gain in a pre-defined grid of possible positions. In the next phase, Alice chooses a trajectory of positions, moves there, and transmits a message to Bob, on the ground, which will estimate the channel gain. The choice of the positions is made both to maximize the mutual information between Alice and Bob and, at the same time, to minimize the information leakage to Eve.

The rest of this work is organized as follows: Section~\ref{sec: sys model} introduces the System model. The proposed key generation protocol is described in Section~\ref{sec:key ext protocol}, and the optimization of the secret-key capacity is in Section~\ref{sec:key cap optimization}. The performance of the proposed solution is evaluated and discussed in Section~\ref{sec: num results}. Lastly, in Section~\ref{sec: conclusions} we draw the main conclusions.

\section{System Model} \label{sec: sys model}

Our scenario consists of a movable device, Alice, and two fixed devices on the ground, namely Bob and Eve. All the devices are equipped with a single antenna. Alice and Bob aim to establish a secret key at the physical layer. On the other hand, Eve, a curious device whose position is known by Alice, wants to infer the key. Alice moves on a grid of discrete positions, each of them associated with a channel gain, and transmits messages to Bob. From each message, Bob measures the channel gain, which is then mapped to a sequence of bits that will constitute the key. For the sake of a simple explanation, we focus on the case where Alice only moves horizontally, always keeping the same altitude $h$. Let us define the set of Alice's positions as  
\begin{equation} \label{eq:ALice_set_pos}
\mathcal{A} = \left\{\bm{p}_{\rm A}(i)=(p_{{\rm A},i,1},p_{{\rm A},i,2},h), \quad i=1,\ldots, M\right\}\,.
\end{equation}
This is the set of $M$ discrete positions on the grid that Alice can pick.

\subsection{Channel Model}

To generate the key, Alice transmits a set of publicly available pilots  $\{b_k\}$ from a position $\bm{p}_{\rm A}\in \mathcal{A}$. 
The communication channel to a receiver in position $\bm{p}_{\rm R}$ is modeled following the Rician fading model \cite[Sec.~2.4]{tse2005fundamentals}. Let $h_k \sim \mathcal{CN}(0, \sigma_h^2)$ be the fading term,  $g(\bm{p}_{\rm A},\bm{p}_{\rm R})$ be the total channel gain, combination of pathloss and shadowing effects, and $\kappa$ the ratio of the energy of the main path to the energy in the scattered paths: the larger it is, the more deterministic is the channel. We fix fading and pilot power to 1, i.e., $\mathbb{E}(\|h_k\|^2)=1$ and $\|b_k\|^2 =1$ for each $k$, where $\mathbb{E}(\cdot)$ is the expectation operator. The corresponding received signal, assuming perfect synchronization with the LOS component, sampled at discrete time $k$  in position $\bm{p}_{\rm R}$ can be written as 
    \begin{equation} \label{eq:rx_signal}
         r_k(\bm{p}_{\rm A},\bm{p}_{\rm R}) = \left( \sqrt{\frac{\kappa}{1+\kappa}} + \sqrt{\frac{1}{1+\kappa}} h_k \right) g(\bm{p}_{\rm A},\bm{p}_{\rm R}) b_k+ w_k \,,
     \end{equation}
where $w_k \sim \mathcal{CN}(0, \sigma_w^2)$ is the thermal noise term.

Let us indicate with $A_\mathrm{Tx}$ and $ A_\mathrm{Rx}$ the gains of the transmitter and receiver antennas, $[f_{\rm c}]_{\si{\mega\hertz }}$ the carrier frequency in \si{\mega\hertz} and $[d]_{\si{\kilo\meter }} = \|\bm{p}_{\rm A}-\bm{p}_{\rm R}\|$ the distance (in \si{\kilo\meter}) between transmitter Alice and the receiver. Then, the path-loss attenuation, according to the Friis model, (in \si{\decibel}) is
\begin{equation}\label{eq:pathloss}
\begin{split}
 [a_{\rm pl}(\bm{p}_{\rm A},\bm{p}_{\rm R})]_{\rm dB}= &32.4+20\log_{10}{[d]_{\si{\kilo\meter }}}+ \\
 & 20\log_{10}{[f_{\rm c}]_{\si{\mega\hertz }}}    -A_\mathrm{Tx}- A_\mathrm{Rx}\,.
\end{split}
\end{equation}
The shadowing effect is described by the Gudmundson model \cite{gudmundson1991correlation}, in which the shadowing attenuation  $[a_{\rm sh}(\bm{p}_{\rm A},\bm{p}_{\rm R})]_{\rm dB}$ (in dB) follows a zero-mean normal distribution with variance $\sigma^{2}_{\rm sh}$, i.e., $[a_{\rm sh}(\bm{p}_{\rm A},\bm{p}_{\rm R})]_{\rm dB} \sim \mathcal{N}(0,\sigma^{2}_{\rm sh})$. Note that the shadowing attenuations are spatially correlated. Indeed, if Alice transmits from positions $\bm{p}_{\rm A}$ and $\bm{p}_{\rm A}'$ the resulting attenuation $[a_{\rm sh}]_{\rm dB} =[a_{\rm sh}(\bm{p}_{\rm A},\bm{p}_{\rm R})]_{\rm dB}$ and $[a_{\rm sh}]'_{\rm dB}=[a_{\rm sh}(\bm{p}_{\rm A}',\bm{p}_{\rm R})]_{\rm dB}$ will have correlation
\begin{equation}\label{eq:gudmundson}
\mathbb{E}\left([a_{\rm sh}]_{\rm dB}[a_{\rm sh}]'_{\rm dB}\right ) =\sigma^{2}_{\rm sh} \exp\left(-\frac{\|\bm{p}_{\rm A} -\bm{p}_{\rm A}'\|}{d_\mathrm{ref}}\right),
\end{equation}
where $d_{\rm ref}$ is the coherence distance \cite{bentom}. This expression shows that the correlation decreases exponentially with the distance between the two positions.

Letting $[a_{\rm tot}(\bm{p}_{\rm A},\bm{p}_{\rm R})]_{\rm dB}=[a_{\rm pl}(\bm{p}_{\rm A},\bm{p}_{\rm R})]_{\rm dB}+[a_{\rm sh}(\bm{p}_{\rm A},\bm{p}_{\rm R})]_{\rm dB}$ be the total channel attenuation, the total channel gain in linear scale $g(\bm{p}_{\rm A},\bm{p}_{\rm R})$ in \eqref{eq:rx_signal} is computed as
\begin{equation} \label{eq:tot:channel}
    g(\bm{p}_{\rm A},\bm{p}_{\rm R}) = 10^{-\frac{[a_{\rm tot}(\bm{p}_{\rm A},\bm{p}_{\rm R})]_{\rm dB}}{20}}\,.
\end{equation}

\subsection{Channel Estimation}

Since the thermal noise term $w_k$ and fading term $h_{k}$ are mutually independent and independent over time, the receiver estimates the channel gain by processing the received signal \eqref{eq:rx_signal} and computes 
\begin{equation} \label{eq:proc_2}
         \tilde{g}'(\bm{p}_{\rm A},\bm{p}_{\rm R}) =  \frac{1}{K}\sum_{k=1}^K r_kb_k^* =\sqrt{\frac{\kappa}{1+\kappa}}g(\bm{p}_{\rm A},\bm{p}_{\rm R}) +  \tilde{w}'_k \,,
     \end{equation}
     where $\tilde{w}'_k \sim \mathcal{CN}(0,\frac{1}{K}(\frac{g(\bm{p}_{\rm A},\bm{p}_{\rm R})^2}{1+\kappa}+\sigma_w^2))$.
     We notice from \eqref{eq:proc_2} that the useful information is purely real, thus by taking the real part, we obtain
        \begin{equation} \label{eq:proc_3}
        \begin{split}
         \tilde{g}(\bm{p}_{\rm A},\bm{p}_{\rm R}) =  \Re(\tilde{g}'(\bm{p}_{\rm A},\bm{p}_{\rm R})) &= \sqrt{\frac{\kappa}{1+\kappa}}g(\bm{p}_{\rm A},\bm{p}_{\rm R}) +  \tilde{w} \\
         &= m(\bm{p}_{\rm A},\bm{p}_{\rm R})+\tilde{w} \,,            
        \end{split}
     \end{equation}
     where $\tilde{w}\sim\mathcal{N}(0,\frac{1}{2K}(\frac{g(\bm{p}_{\rm A},\bm{p}_{\rm R})^2}{1+\kappa}+\sigma_w^2))$, and we defined $m(\bm{p}_{\rm A},\bm{p}_{\rm R}) =\sqrt{\frac{\kappa}{1+\kappa}}g(\bm{p}_{\rm A},\bm{p}_{\rm R})$ as the expected gain at the receiver.

\subsection{Knowledge Assumptions}

\paragraph*{Alice Knowledge}
The information that Alice has about the curious device Eve are: a) Eve's location $\bm{p}_{\rm E}$, and b) the function $m(\bm{p}_{\rm A},\bm{p}_{\rm E})$ for each tuple $(\bm{p}_{\rm A},\bm{p}_{\rm E})$.

\paragraph*{Eve Knowledge}
The curious device Eve, located in position $\bm{p}_{\rm E}$, knows:
\begin{itemize}
    \item The set of possible Alice position's $\mathcal{A}$;
    \item  The function $m(\bm{p}_{\rm A},\bm{p}_{\rm E})$ for each tuple $(\bm{p}_{\rm A},\bm{p}_{\rm E})$, that we denote in the following as {\em map};
    \item  The pilot sequences $\{b_k\}$ used by both Alice and Bob.
\end{itemize}
Eve, upon the reception of the transmitted pilot signals, estimates the gain using \eqref{eq:proc_3} and obtains the sequence 
$\bm{t}=\{\tilde{g}(\bm{p}_{{\rm A},n},\bm{p}_{\rm E})\}$.
      
\section{Key Extraction Protocol} \label{sec:key ext protocol}
We assume Bob and Eve are located in fixed and known positions $\bm{p}_{\rm B}$ and $\bm{p}_{\rm E}$, Alice, on the other hand, will be free to move in $\mathcal{A}$. Alice uses her movements to generate the key from the channel gain. The reason for exploiting the movements rather than varying the transmission power and/or exploiting phase information is twofold. First, power information is available in most off-the-shelf devices and is more robust to phase synchronization issues \cite{shehadeh2015survey}. Second, simply varying the transmitting power would leak information to the curious device, as each variation would be detected by both Bob and Eve. On the other hand, by exploiting the transmitting positions with the drone mobility, we can completely masquerade the power variations at Eve, thus achieving good security performance as shown in Sections~\ref{sec:key cap optimization} and~\ref{sec: num results}.

The key generation protocol is outlined as follows:
\begin{enumerate}
\item \emph{Training Phase}: Alice visits all the positions in $\mathcal{A}$ and in each position Bob transmits pilot signals that enable Alice to estimate the channel gain between her and Bob. From~\eqref{eq:proc_3}, Alice estimates the function mapping Alice and Bob positions to the gain between them $m(\bm{p}_{\rm A},\bm{p}_{\rm B})$ for each pair $(\bm{p}_{\rm A},\bm{p}_{\rm B})$. We refer as $\mathcal{M}=\{m(\bm{p}_{\rm A},\bm{p}_{\rm B}), \forall \bm{p}_{\rm A}\}$ the set of possible gains. Similarly, by letting Eve transmit pilots in correspondence of each position assumed by Alice, Alice can estimate the channel gain between her and the curious device Eve (in position $\bm{p}_{\rm E}$) and obtains the map $m(\bm{p}_{\rm A},\bm{p}_{\rm E})$. We recall that Eve is a curious receiver who obeys the rules of the network and thus will send pilots when instructed to do so. Note that as Alice will transmit pilots from her positions (see step 2)), also the curious device can estimate the mapping $m(\bm{p}_{\rm A},\bm{p}_{\rm E})$, thus Alice has no advantage in this phase over Eve. 
\item \emph{Random Positions Choice}: Alice computes a probability distribution $P_{\bm p_{\rm A}}$ over the set of positions $\mathcal{A}$ and extracts independently a set of $N$ positions  $\bm{P}=\{\bm{p}_{{\rm A},n} \sim P_{\bm p_{\rm A}}, n=1, \ldots, N\}$. Next, Alice moves to those positions and transmits a pilot sequence $\{b_k\}$ of $K$ symbols.
\item  \emph{Gain Estimation}: Bob, upon the reception of the message, estimates the gain using \eqref{eq:proc_3}. After $N$ transmissions, Bob will have obtained a sequence of $N$ estimated gains $\tilde{\bm{g}} = \{\tilde{g}(\bm{p}_{{\rm A},n},\bm{p}_{\rm B}), n=1, \ldots, N\}$.
\item \emph{Key Distillation}:
Bob quantizes the received gains with the quantization step $\Delta$. Letting $\mathcal{Q}=\{q_1,\ldots,q_Q\}$ be the quantization levels, at each transmission $n$ Bob obtains
\begin{equation}
    q_{\bm{p}_{{\rm A},n}}=Q(\tilde{g}(\bm{p}_{{\rm A},n},\bm{p}_{\rm B})) \in \mathcal{Q}\, ,
\end{equation}
where $Q(\cdot)$ is the quantization function. Note that, as the quantization can be made arbitrarily small (at the expense of optimization time and memory consumption), we will study its impact in Section~\ref{sec: num results}. By mapping each of these levels to a unique sequence of bits, Bob distills the key $\bm{q}=\{q_{\bm{p}_{{\rm A},n}}, n=1, \ldots, N\}$. Alice, from \eqref{eq:proc_3}, has the sequence of expected gains  $\bm{m} = \{m(\bm{p}_{{\rm A},n},\bm{p}_{\rm B}), n=1, \ldots, N\}$.  
\item \emph{Information Reconciliation}: The two sequences obtained in the previous point will have a high correlation as we can think of $\bm{q}$ as the noisy version of $\bm{m}$, due to quantization and thermal noises. Thus, information reconciliation is needed to agree on a common key.
\item \emph{Privacy Amplification}:  Alice and Bob agree on a public hashing function to apply on the reconciled keys; the purpose is to reduce the size of the key to the number of bits effectively secret to Eve. Yet, as we demonstrate in the next Section~\ref{sec:security_analysis_finale}, it is not needed in our protocol. 
\end{enumerate}

\subsection{Security Analysis} \label{sec:security_analysis_finale}

The proposed protocol falls under the {\em channel-based \ac{skg}} mechanisms \cite[Chp.~4]{bloch2008physical}, as Alice has control of the channel by picking a sequence of positions $\bm{P}$, that will result in keys $\bm{q}$, $\bm{m}$, and $\bm{t}$ at Bob, Alice, and Eve respectively. Since Alice picks the positions independently, we can study the security of the protocol at the general $n$-th transmission, and then the effects of multiple transmissions are simple to analyze. 
Let us fix the number of transmissions to $N=1$, and let $m_{\bm{p}_{{\rm A}}} = m(\bm{p}_{{\rm A},1},\bm{p}_{\rm B})$, $q_{\bm{p}_{{\rm A}}} = Q(\tilde{g}(\bm{p}_{{\rm A},1},\bm{p}_{\rm B}))$ and $t_{\bm{p}_{{\rm A}}} = \tilde{g}(\bm{p}_{{\rm A},1},\bm{p}_{\rm E})$ be the resulting keys. 

In this paper, the performance metric is the \ac{skc}, i.e., an upper bound on the number of secret bits that the legitimate parties can share at each transmission. More precisely, we will maximize a lower bound on the \ac{skc} (in bits per position)  \cite[Sec.~4.4]{bloch2008physical}
\begin{equation} \label{eq:sec_key_rate}
\begin{split}
    \mathcal{C}_{\rm key} = \max [ & \max_{P_{\bm p_{\rm A}}} \left(\mathbb{I}(q_{\bm{p}_{{\rm A}}};m_{\bm{p}_{{\rm A}}}) - \mathbb{I}(m_{\bm{p}_{{\rm A}}};t_{\bm{p}_{{\rm A}}}\right), \\
  &\max_{P_{\bm p_{\rm A}}} \left(\mathbb{I}(q_{\bm{p}_{{\rm A}}};m_{\bm{p}_{{\rm A}}}) - \mathbb{I}(q_{\bm{p}_{{\rm A}}};t_{\bm{p}_{{\rm A}}}\right)] \,,
\end{split} 
\end{equation}
where $\mathbb{I}(\cdot;\cdot)$ is the mutual information operator.

Here we introduce the concept of \emph{isohypse} that partitions the position set $\mathcal{A}$ into $L$ subsets
    \begin{equation}
        \mathcal{A} = \{\mathcal I_1,\ldots, \mathcal I_L\}\,,
    \end{equation}
where isohypse $\mathcal I_{\ell}= \{\bm{p}_{\rm A} \in \mathcal A: m(\bm{p}_{\rm A},\bm{p}_{\rm E})=m_{\ell}\} $ contains the Alice's positions leading the \emph{same} (expected) gain $m_{\ell}$ at Eve. This partition allows for simplifying the analysis; in fact, if we restrict the movement of Alice to an isohypse ${\mathcal I}_{\ell}$, Eve's measurements will be independent of Alice and Bob. This is because Eve will be measuring a constant value $m_{\ell}$ affected by noise leading $\mathbb{I}(m_{{\bm p}_{\mathrm{A}}};t_{\bm{p}_{{\mathrm{A}}}}) =0$ and $\mathbb{I}(q_{\bm{p}_{{\mathrm{A}}}};t_{\bm{p}_{{\mathrm{A}}}}) =0$  in \eqref{eq:sec_key_rate}. Thus, if Alice moves along an isohypse, no leakage of the secret key is given to Eve. Therefore, step 6 of privacy amplification is not needed in our proposed protocol.

\subsection{Attack Strategy}
It is clear from the previous subsection that the attacker has no strategies available, as the isohypse movement restriction leads to an information leakage of zero, regardless of its position on the ground. Yet, depending on the proximity of Eve to Bob, the number of secret bits Bob can extract may vary, as will be shown in Section~\ref{sec: num results}.

\section{Secret-key Capacity Optimization} \label{sec:key cap optimization}
If Alice moves on the isohypse ${\mathcal I}_{\ell}$, we write the obtained \ac{skc} following  \eqref{eq:sec_key_rate} as
\begin{equation} \label{eq:sec_key_rate_1}
\mathcal{C}_{\rm key}({\mathcal I}_{\ell})=\mathbb{I}(q_{\bm{p}_{{\rm A}}};m_{\bm{p}_{{\rm A}}}) = H(q_{\bm{p}_{{\rm A}}})-H(q_{\bm{p}_{{\rm A}}}|m_{\bm{p}_{{\rm A}}})\,,
\end{equation}
where $H(\cdot)$ is the entropy operator.  Let $ P_{\bm p_{\rm A}|{\mathcal I}_\ell}(\bm{p})$ the position probability distribution of Alice $P_{\bm p_{\rm A}}$ conditioned on the movement on the isohypse ${\mathcal I}_{\ell}$. We compute the \ac{pmd} of the obtained quantized gains $ P_{q_{\bm{p}_{{\rm A}}}}(q)$ as
    \begin{equation}
        P_{q_{\bm{p}_{{\rm A}}}}(q)=\sum_{\bm{p} \in {\mathcal I}_{\ell}}P_{q_{\bm{p}_{{\rm A}}}}(q|\bm{p}_{\rm A}=\bm{p})P_{\bm p_{\rm A}|{\mathcal I}_\ell}(\bm p)\,.
    \end{equation}
We define $g(\cdot;\bm{p}_{\rm A})$ as the Gaussian PDF centered in $m_{\bm{p}_{\rm A}}$ and variance $\sigma^2 = \frac{1}{2K}(\frac{g(\bm{p}_{\rm A},\bm{p}_{\rm B})^2}{1+\kappa}+\sigma_w^2)$, then using \eqref{eq:proc_3} we have
\begin{equation} \label{eq:q_given_p}
    P_{q_{\bm{p}_{{\rm A}}}}(q|\bm{p}_{\rm A} = \bm{p})= \int_{-\frac{\Delta}{2}}^{\frac{\Delta}{2}} g(q+a;\bm{p}) \,da = {\rm Q}'_+ - {\rm Q}'_-\,,
\end{equation}
where ${\rm Q}'_{\pm} = {\rm Q_F}[(q  \pm\Delta/2- m_{\bm{p}_{\rm A}})/\sigma]$ and ${\rm Q_F}(\cdot)$ is the ${\rm Q}$-function. Thus by definition
    \begin{equation} \label{eq:entropy_q_pa}
        \begin{split}
            H(q_{\bm{p}_{{\rm A}}}) &= -\sum_{q \in \mathcal{Q}}  P_{q_{\bm{p}_{{\rm A}}}}(q) \log_2P_{q_{\bm{p}_{{\rm A}}}}(q)   \\
            &= -\sum_{q \in \mathcal{Q}}  \sum_{\bm{p} \in {\mathcal I}_{\ell}}P_{q_{\bm{p}_{{\rm A}}}}(q|\bm{p}_{\rm A}=\bm{p})P_{\bm p_{\rm A}|{\mathcal I}_\ell}(\bm p) \times \\
            & \times \log_2 \left(\sum_{\bm{p} \in {\mathcal I}_{\ell}}P_{q_{\bm{p}_{{\rm A}}}}(q|\bm{p}_{\rm A}=\bm{p})P_{\bm p_{\rm A}|{\mathcal I}_\ell}(\bm p) \right).
        \end{split}
    \end{equation}
About the conditioned entropy term $H(q_{\bm{p}_{{\rm A}}}|m_{\bm{p}_{{\rm A}}})$ we have
    \begin{equation} \label{eq:entropy_noise}
        H(q_{\bm{p}_{{\rm A}}}|m_{\bm{p}_{{\rm A}}}) =\hspace{-0.5cm}\sum_{m_{\bm p} \in \mathcal{M}: \bm{p}\in {\mathcal I}_{\ell}  }  H(q_{\bm{p}_{{\rm A}}}|m_{\bm{p}_{{\rm A}}} =m_{\bm p})\mathbb{P}(m_{\bm{p}_{{\rm A}}} =m_{\bm p}).
    \end{equation}
Still, since Eve is sufficiently far from Bob (at least $d_{\rm ref}$  \si{\meter}), and due to the shadowing, typically the mapping between positions in ${\mathcal I}_{\ell}$ and expected gains \emph{at Bob} is unique, i.e.,  $\mathbb{P}(m_{\bm{p}_{{\rm A}}} =m) = P_{\bm p_{\rm A}|{\mathcal I}_\ell}(\bm p_{\rm A})$.
Thus by recalling \eqref{eq:q_given_p}, we have that \eqref{eq:entropy_noise} simplifies into
\begin{equation} \label{eq:q_given_p_2}
    \begin{split}
        H(q_{\bm{p}_{{\rm A}}}|m_{\bm{p}_{{\rm A}}}) &=\sum_{m_{\bm p} \in \mathcal{M}: \bm{p}\in {\mathcal I}_{\ell}  }  H(q_{\bm{p}_{{\rm A}}}|m_{\bm{p}_{{\rm A}}} =m_{\bm p})\mathbb{P}(m_{\bm{p}_{{\rm A}}} =m_{\bm p})\\ 
        &=\sum_{\bm{p} \in {\mathcal I}_{\ell}}  H(q_{\bm{p}_{{\rm A}}}|\bm{p}_{{\rm A}} =\bm{p})P_{\bm p_{\rm A}|{\mathcal I}_\ell}(\bm p) \\
        &=-\sum_{\bm{p} \in {\mathcal I}_{\ell}} \sum_{q \in \mathcal{Q}} P_{\bm p_{\rm A}|{\mathcal I}_\ell}(\bm p) P_{q_{\bm{p}_{{\rm A}}}}(q|\bm{p}_{\rm A} = \bm{p})\times \\
        &\times \log_2P_{q_{\bm{p}_{{\rm A}}}}(q|\bm{p}_{\rm A} = \bm{p})\,. 
    \end{split}
    \end{equation}
To find the optimal distribution $P_{\bm p_{\rm A}}^\star$ maximizing the secrecy capacity \eqref{eq:sec_key_rate} we look for the isohypse leading the highest $\mathcal{C}_{\rm key}({\mathcal I}_{\ell})$, thus 
\begin{equation} \label{eq:opt_problem}
\begin{aligned}
P_{\bm p_{\rm A}}^\star =\arg \max_{\ell}\max_{P_{\bm p_{\rm A}|{\mathcal I}_\ell}(\bm{p})} \quad &   \mathcal{C}_{\rm key}({\mathcal I}_{\ell}) \\
\textrm{s.t.} 
       \quad & P_{\bm p_{\rm A}|{\mathcal I}_\ell}(\bm{p}) \in [0,1]\,\, \forall \bm{p}, \ell\\
        & \sum_{\bm{p}\in {\mathcal I}_{\ell}}P_{\bm p_{\rm A}|{\mathcal I}_\ell}(\bm{p})=1\,\, \forall \ell\,.
\end{aligned}
\end{equation}

We note from \eqref{eq:q_given_p_2} that in the very low-noise and dense quantization case,  for a given $\bm{p}$, $P_{q_{\bm{p}_{{\rm A}}}}(q|\bm{p}_{\rm A} = \bm{p})=1$ for only a specific quantization level $q$ and zero otherwise. This yields to the conditional entropy in \eqref{eq:q_given_p_2} equal to zero and \eqref{eq:entropy_q_pa} provides $ H(q_{\bm{p}_{{\rm A}}})= -\sum_{\bm{p} \in {\mathcal I}_{\ell}}P_{\bm p_{\rm A}|{\mathcal I}_\ell}(\bm p) \log_2 P_{\bm p_{\rm A}|{\mathcal I}_\ell}(\bm p)$. Consequently, the problem can be directly mapped to the optimal choice over a finite alphabet ${\mathcal I}_{\ell}$. The optimal distribution is then uniform over the alphabet, i.e., $P_{\bm p_{\rm A}|{\mathcal I}_\ell}(\bm{p}) =\frac{1}{|{\mathcal I}_{\ell}|}$ for each $\bm{p} \in \mathcal I_\ell$. Lastly, we can bound \eqref{eq:sec_key_rate_1} with $\log_2|{\mathcal I}_{\ell}|$ obtaining the following upper bound to the secret key capacity 
\begin{equation} \label{eq:upper_bound_tot}
    \mathcal{C}_{\rm key} = \max_{\ell}\mathcal{C}_{\rm key}(\mathcal I_{\ell})  \leq \bar{\mathcal{C}}_{\rm key} =\max_{\ell}\log_2|{\mathcal I}_{\ell}|.
\end{equation}
\section{Numerical Results} \label{sec: num results}
We numerically simulated the proposed protocol to evaluate the \ac{skc} in terms of the number of secret key bits per position as a performance metric. First, we define parameters. From \eqref{eq:tot:channel}, we define the maximum gain expected for Bob as $
    g_{\rm max} = \max_{\bm{p}_{\rm A}}g(\bm{p}_{\rm A},\bm{p}_{\rm B})$,
and we measure the tradeoff between pathloss-shadowing and fading components in \eqref{eq:rx_signal} by the parameter (in dB)
\begin{equation}
    \rho_{\rm dB} = 10\log_{10}\frac{\kappa}{g_{\rm max}}\,.
\end{equation}
High values of $\rho_{\rm dB}$ are expected, as in this paper, we deal with poor scattering environments. 
When Alice is in position $\bm{p}_{\rm A}$, the \ac{snr} at Bob from \eqref{eq:proc_3} is
     \begin{equation}
         {\rm SNR}_{\bm{p}_{\rm A}} = \frac{g(\bm{p}_{\rm A},\bm{p}_{\rm B})^2   \frac{\kappa}{1+\kappa}}{ \frac{1}{2}(\frac{g(\bm{p}_{\rm A},\bm{p}_{\rm B})^2}{1+\kappa}+\sigma_w^2)}=\frac{2g(\bm{p}_{\rm A},\bm{p}_{\rm B})^2 \kappa}{g(\bm{p}_{\rm A},\bm{p}_{\rm B})^2 + \sigma_w^2(1+\kappa)}\,,
     \end{equation}
     and the minimum ${\rm SNR}$ of operation is $         {\rm SNR}_{\rm min} = \min_{\bm{p}_{\rm A} \in \mathcal{A}} {\rm SNR}_{\bm{p}_{\rm A}}\,$. 
The shadowing \ac{std} (see \eqref{eq:gudmundson}) for drone communications is given in~\cite{sharma2018study} for devices transmitting at $f_{\rm c}=2.4$ \si{\giga \hertz}. Alice moves on a squared grid of dimension $500 \si{\meter} \times 500 \si{\meter}$ at altitude $h=10$ \si{\meter}. Unless otherwise stated, we consider: $\sigma_{\rm sh,E} = 3$ \si{\decibel} as the shadowing variance of Alice-Eve channel, $\sigma_{\rm sh,A} = 5$ \si{\decibel} as the shadowing variance of the Alice-Bob channel, a $d = 119$ \si{\meter} distance between Eve and Bob, $K=10$, ${\rm SNR}_{\rm min}=10$ \si{\decibel}, $\rho_{\rm dB} = 10$ \si{\decibel}, $Q=16$, and $M=150 \times 150$. To solve \eqref{eq:opt_problem}, we used the Sequential Least Squares Programming method.

\subsection{Map Geometry and Number of Quantization Levels}
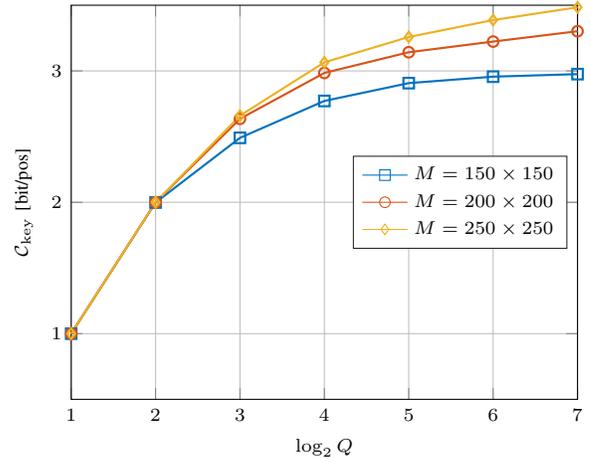
\begin{figure}
    \centering
        \input{figures/fig_ptax_quant}
\caption{\ac{skc} vs the number of quantization levels $Q$ (in log scale), with different map sampling spaces.} 
    \label{fig: ptax_quant}
\end{figure}

We see from Fig. \ref{fig: ptax_quant} that by increasing the number of quantization levels $Q$, the secret key capacity first increases and then saturates. This justifies our design decision to quantize the estimated gains at Bob and work with discrete levels. The same is true for the number of map positions $M$: a higher number of points on the map improves the performance, yet oversampling the space would result in sampling the same gains, leading to a saturation of the key entropy at Bob.

\subsection{Thermal and Fading Noises, and Number of Pilot Symbols}
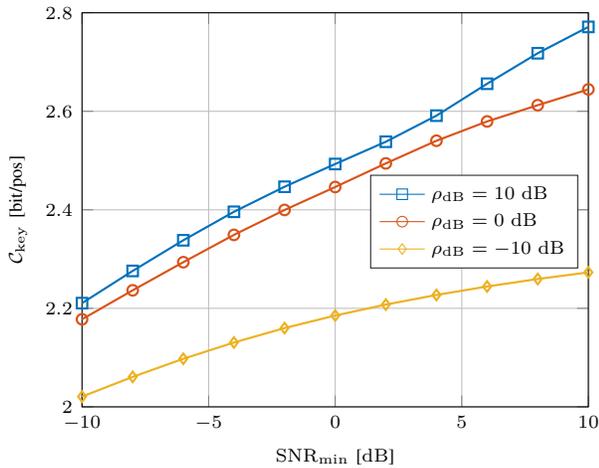
\begin{figure}
    \centering
        \input{figures/fig_snr_k}
\caption{\ac{skc} vs ${\rm SNR}_{\rm min}$, and  pathloss-shadowing and fading tradeoff parameter $\rho_{\rm dB}$.} 
    \label{fig: capacity_snr}
\end{figure}

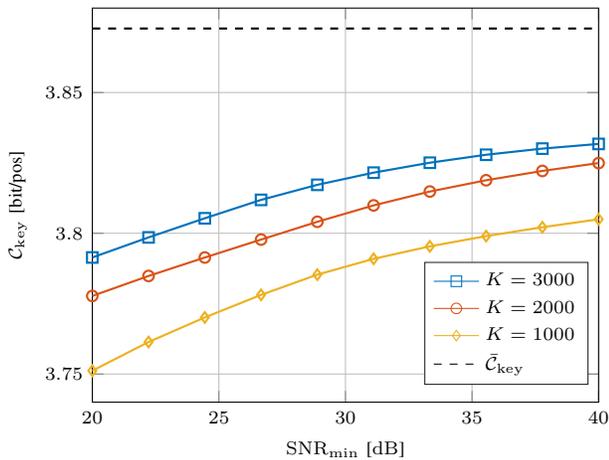
\begin{figure}
    \centering
        \input{figures/fig_snr_k_high}
\caption{\ac{skc} vs ${\rm SNR}_{\rm min}$, the number of pilot symbols $K$, and upper bound $\bar{\mathcal{C}}_{\rm key}$.}
    \label{fig: capacity_snr_high}
\end{figure}

Fig. \ref{fig: capacity_snr} shows that the capacity increases as the thermal noise reduces, as expected by varying ${\rm SNR}_{\rm min}$. Moreover, we showed that the higher $\rho_{\rm dB}$ (which corresponds to less fading) increases the capacity. This is because, from \eqref{eq:proc_3}, low values of $\rho_{\rm dB}$ result in an estimation noise that depends on the realized gain, and this effect worsens the performance. On the other hand, Fig. \ref{fig: capacity_snr_high} demonstrates the limits of the protocol, at very high \ac{snr} and number of pilots in comparison with the bound, obtainable with infinite dense quantization and zero noise. Here we used $Q=128$ quantization levels. We see that, even with high but reasonable simulation parameters, the performance is not far from the best achievable.

\subsection{Shadowing Variance and Eve-Bob Distance}
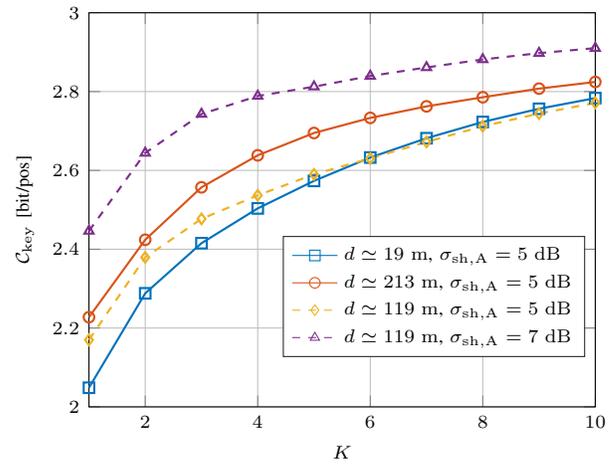
\begin{figure}
    \centering
        \input{figures/fig_sigma_dist}
\caption{\ac{skc} as a function of the number of pilot symbols $K$, with different Alice-Bob channel shadowing \ac{std} $\sigma_{\rm sh,A}$ and Eve-Bob distance $d$.} 
    \label{fig: sigma_dist}
\end{figure}

Fig. \ref{fig: sigma_dist} shows that higher shadowing \ac{std} at Bob leads to higher capacity, this is because given an isohypse, the resulting gains at Bob have higher entropy. Finally, the proximity of Eve to Bob worsens the performance because all the isohypses lead to similar gains to Bob: if Eve were exactly in Bob's position, all isohypses by definition would lead to the same gain also to Bob, and the mutual information between him and Alice would be zero as well.

\balance

\section{Conclusions} \label{sec: conclusions}
In this paper, we proposed a novel solution that exploits the mobility of the \acp{uav} to perform \ac{skg} at the physical layer in the presence of a curious device, over Rician fading channels.  The effectiveness of our protocol was demonstrated via numerical simulations, showing that it is possible to obtain more than 2 bits per position of the secret key even in poor scattering and slow fading environments.
\bibliographystyle{IEEEtran}
\bibliography{IEEEabrv,biblio}

\end{document}

%% file: myacro.tex
\newacronym{awgn}{AWGN}{additive white Gaussian noise}

\newacronym{cdf}{CDF}{cumulative distribution function}

\newacronym{mimo}{MIMO}{multiple-input and multiple-output}
\newacronym{ml}{ML}{maximum likelihood}

\newacronym[longplural=Nash equilibria]{ne}{NE}{Nash equilibrium}

\newacronym{pdf}{pdf}{probability density function}
\newacronym{psk}{PSK}{phase-shift keying}
\newacronym{nlos}{NLOS}{not line-of-sight}
\newacronym{skc}{SKC}{secret-key capacity}
\newacronym{pmd}{PMD}{probability mass distribution}
\newacronym{los}{LOS}{line-of-sight}
\newacronym{skg}{SKG}{secret-key generation}
\newacronym{uav}{UAV}{unmanned aerial device}

\newacronym{snr}{SNR}{signal-to-noise ratio}
\newacronym{std}{STD}{standard deviation}

%% file: figures/fig_ptax_quant.tex
%
\definecolor{mycolor1}{rgb}{0.00000,0.44700,0.74100}%
\definecolor{mycolor2}{rgb}{0.85000,0.32500,0.09800}%
\definecolor{mycolor3}{rgb}{0.92900,0.69400,0.12500}%
\begin{tikzpicture}[ every plot/.style={thick}]

\begin{axis}[%
width=0.95\fwidth,
height=0.74\fheight,
at={(0\fwidth,0\fheight)},
scale only axis,
xmin=1,
xmax=7,
xlabel style={font=\color{white!15!black}},
xlabel={$\log_2 Q$},
ymin=0.5,
ymax=3.5,
ylabel style={font=\color{white!15!black}},
ylabel={$\mathcal{C}_{\rm key}$ [bit/pos]},
axis background/.style={fill=white},
xmajorgrids,
ymajorgrids,
legend style={at={(0.98,0.5)}, anchor=east, legend cell align=left, align=left, draw=white!15!black},
enlargelimits=false,title style={font=\scriptsize},xlabel style={font=\scriptsize},ylabel style={font=\scriptsize},legend style={font=\scriptsize},ticklabel style={font=\scriptsize}
]
\addplot [color=mycolor1, mark=square, mark options={solid, mycolor1}]
  table[row sep=crcr]{%
1	0.999999999063853\\
2	1.9981943046575\\
3	2.49052825587976\\
4	2.77130085034949\\
5	2.90707424818099\\
6	2.95619116466046\\
7	2.97558051110743\\
};
\addlegendentry{$M = 150 \times 150 $}

\addplot [color=mycolor2, mark=o, mark options={solid, mycolor2}]
  table[row sep=crcr]{%
1	0.999999999999983\\
2	1.99997440493476\\
3	2.63530240807803\\
4	2.98447204925096\\
5	3.14209648936062\\
6	3.22343413920368\\
7	3.30209742377643\\
};
\addlegendentry{$M = 200 \times 200 $}

\addplot [color=mycolor3, mark=diamond, mark options={solid, mycolor3}]
  table[row sep=crcr]{%
1	0.999999999999976\\
2	1.99998940464669\\
3	2.6592625107784\\
4	3.06515897164832\\
5	3.25822816918685\\
6	3.38704258269725\\
7	3.48434847667366\\
};
\addlegendentry{$M = 250 \times 250 $}

\end{axis}
\end{tikzpicture}%

%% file: figures/fig_snr_k.tex
%
\definecolor{mycolor1}{rgb}{0.00000,0.44700,0.74100}%
\definecolor{mycolor2}{rgb}{0.85000,0.32500,0.09800}%
\definecolor{mycolor3}{rgb}{0.92900,0.69400,0.12500}%
\begin{tikzpicture}[ every plot/.style={thick}]

\begin{axis}[%
width=0.95\fwidth,
height=0.74\fheight,
at={(0\fwidth,0\fheight)},
scale only axis,
xmin=-10,
xmax=10,
xlabel style={font=\color{white!15!black}},
xlabel={$\rm SNR_{min}$ [\si{\decibel}] },
ymin=2,
ymax=2.8,
ylabel style={font=\color{white!15!black}},
ylabel={$\mathcal{C}_{\rm key}$ [bit/pos]},
axis background/.style={fill=white},
xmajorgrids,
ymajorgrids,
legend style={at={(0.98,0.47)}, anchor=east, legend cell align=left, align=left, draw=white!15!black},
enlargelimits=false,title style={font=\scriptsize},xlabel style={font=\scriptsize},ylabel style={font=\scriptsize},legend style={font=\scriptsize},ticklabel style={font=\scriptsize}
]
\addplot [color=mycolor1, mark=square, mark options={solid, mycolor1}]
  table[row sep=crcr]{%
-10	2.21057936726096\\
-8	2.27593161685705\\
-6	2.33814980755395\\
-4	2.39595614697086\\
-2	2.44682228957187\\
0	2.4929813968423\\
2	2.53816257153363\\
4	2.59115464557413\\
6	2.65591385727216\\
8	2.71785057991518\\
10	2.77129439046491\\
};
\addlegendentry{$\rho_{\rm dB}= 10 $ \si{\decibel}}

\addplot [color=mycolor2, mark=o, mark options={solid, mycolor2}]
  table[row sep=crcr]{%
-10	2.17774941771802\\
-8	2.23654858383543\\
-6	2.29365853624743\\
-4	2.34907645678893\\
-2	2.39970391357619\\
0	2.4462858252581\\
2	2.49423997280688\\
4	2.54021061901479\\
6	2.57939087822615\\
8	2.61237105607029\\
10	2.64445350462572\\
};
\addlegendentry{$\rho_{\rm dB}= 0 $ \si{\decibel}}

\addplot [color=mycolor3, mark=diamond, mark options={solid, mycolor3}]
  table[row sep=crcr]{%
-10	2.02082030136865\\
-8	2.06081562242889\\
-6	2.097541583924\\
-4	2.13063182220171\\
-2	2.15988941611678\\
0	2.1853344075722\\
2	2.20759961319699\\
4	2.22714476356624\\
6	2.24433643465986\\
8	2.2595168839938\\
10	2.27298704630484\\
};
\addlegendentry{$\rho_{\rm dB}= -10 $ \si{\decibel}}

\end{axis}
\end{tikzpicture}%

%% file: figures/fig_snr_k_high.tex
%
\definecolor{mycolor1}{rgb}{0.00000,0.44700,0.74100}%
\definecolor{mycolor2}{rgb}{0.85000,0.32500,0.09800}%
\definecolor{mycolor3}{rgb}{0.92900,0.69400,0.12500}%
\begin{tikzpicture}[ every plot/.style={thick}]

\begin{axis}[%
width=0.95\fwidth,
height=0.74\fheight,
at={(0\fwidth,0\fheight)},
scale only axis,
xmin=20,
xmax=40,
xlabel style={font=\color{white!15!black}},
xlabel={$\rm SNR_{min}$ [\si{\decibel}]},
ymin=3.74,
ymax=3.88,
ylabel style={font=\color{white!15!black}},
ylabel={$\mathcal{C}_{\rm key}$ [bit/pos]},
axis background/.style={fill=white},
xmajorgrids,
ymajorgrids,
legend style={at={(0.98,0.2)}, anchor=east, legend cell align=left, align=left, draw=white!15!black},
enlargelimits=false,title style={font=\scriptsize},xlabel style={font=\scriptsize},ylabel style={font=\scriptsize},legend style={font=\scriptsize},ticklabel style={font=\scriptsize}
]
\addplot [color=mycolor1, mark=square, mark options={solid, mycolor1}]
  table[row sep=crcr]{%
20	3.79140110487973\\
22.2222222222222	3.79853280821277\\
24.4444444444444	3.80532702827693\\
26.6666666666667	3.81185985808327\\
28.8888888888889	3.81721473980838\\
31.1111111111111	3.82153423226694\\
33.3333333333333	3.82506335847\\
35.5555555555556	3.82789249265909\\
37.7777777777778	3.83007508460518\\
40	3.83173868506179\\
};
\addlegendentry{$K= 3000$}

\addplot [color=mycolor2, mark=o, mark options={solid, mycolor2}]
  table[row sep=crcr]{%
20	3.77773526301293\\
22.2222222222222	3.78483659068598\\
24.4444444444444	3.79138711364666\\
26.6666666666667	3.79776288319281\\
28.8888888888889	3.80407777587718\\
31.1111111111111	3.80990308043425\\
33.3333333333333	3.814826921416\\
35.5555555555556	3.81883230371305\\
37.7777777777778	3.82214189822226\\
40	3.82493111061614\\
};
\addlegendentry{$K= 2000$}

\addplot [color=mycolor3, mark=diamond, mark options={solid, mycolor3}]
  table[row sep=crcr]{%
20	3.75116679038404\\
22.2222222222222	3.76136823491495\\
24.4444444444444	3.77009885199845\\
26.6666666666667	3.77812032421612\\
28.8888888888889	3.78530846275043\\
31.1111111111111	3.79091957947033\\
33.3333333333333	3.79535268355387\\
35.5555555555556	3.79899403866479\\
37.7777777777778	3.80214360148715\\
40	3.80497020749946\\
};
\addlegendentry{$K= 1000$}

\addplot [color=black, dashed]
  table[row sep=crcr]{%
20	3.87274104263583\\
22.2222222222222	3.87274104263583\\
24.4444444444444	3.87274104263583\\
26.6666666666667	3.87274104263583\\
28.8888888888889	3.87274104263583\\
31.1111111111111	3.87274104263583\\
33.3333333333333	3.87274104263583\\
35.5555555555556	3.87274104263583\\
37.7777777777778	3.87274104263583\\
40	3.87274104263583\\
};
\addlegendentry{$\bar{\mathcal{C}}_{\rm key}$}

\end{axis}
\end{tikzpicture}%

%% file: figures/fig_sigma_dist.tex
%
\definecolor{mycolor1}{rgb}{0.00000,0.44700,0.74100}%
\definecolor{mycolor2}{rgb}{0.85000,0.32500,0.09800}%
\definecolor{mycolor3}{rgb}{0.92900,0.69400,0.12500}%
\definecolor{mycolor4}{rgb}{0.49400,0.18400,0.55600}%
\begin{tikzpicture}[ every plot/.style={thick}]

\begin{axis}[%
width=0.95\fwidth,
height=0.74\fheight,
at={(0\fwidth,0\fheight)},
scale only axis,
xmin=1,
xmax=10,
xlabel style={font=\color{white!15!black}},
xlabel={$K$},
ymin=2,
ymax=3,
ylabel style={font=\color{white!15!black}},
ylabel={$\mathcal{C}_{\rm key}$ [bit/pos]},
axis background/.style={fill=white},
xmajorgrids,
ymajorgrids,
legend style={at={(0.98,0.28)}, anchor=east, legend cell align=left, align=left, draw=white!15!black},
enlargelimits=false,title style={font=\scriptsize},xlabel style={font=\scriptsize},ylabel style={font=\scriptsize},legend style={font=\scriptsize},ticklabel style={font=\scriptsize}
]
\addplot [color=mycolor1, mark=square, mark options={solid, mycolor1}]
  table[row sep=crcr]{%
1	2.04855193739736\\
2	2.28805831435187\\
3	2.41533456639737\\
4	2.50360438761947\\
5	2.57339337952881\\
6	2.6326567732536\\
7	2.68164895889824\\
8	2.72290705947836\\
9	2.75637871404182\\
10	2.78348262617487\\
};
\addlegendentry{$d \simeq 19 $ \si{\meter}, $\sigma_{\rm sh,A}= 5$ \si{\decibel}}

\addplot [color=mycolor2, mark=o, mark options={solid, mycolor2}]
  table[row sep=crcr]{%
1	2.22755341906904\\
2	2.42396733021429\\
3	2.55708462319661\\
4	2.63827667010698\\
5	2.69491615346191\\
6	2.7332579401018\\
7	2.76248356576574\\
8	2.78563631297654\\
9	2.80749496701075\\
10	2.82445708737811\\
};
\addlegendentry{$d \simeq 213$ \si{\meter}, $\sigma_{\rm sh,A}= 5$ \si{\decibel}}

\addplot [color=mycolor3, dashed, mark=diamond, mark options={solid, mycolor3}]
  table[row sep=crcr]{%
1	2.1700801192049\\
2	2.37927924848875\\
3	2.47697944668398\\
4	2.53675466018203\\
5	2.59003208948594\\
6	2.63044506548569\\
7	2.67233788229939\\
8	2.71192660158219\\
9	2.74419460800751\\
10	2.77129146823585\\
};
\addlegendentry{$d \simeq 119$ \si{\meter}, $\sigma_{\rm sh,A}= 5$ \si{\decibel}}

\addplot [color=mycolor4, dashed, mark=triangle, mark options={solid, mycolor4}]
  table[row sep=crcr]{%
1	2.446295259202\\
2	2.64418270176892\\
3	2.74293941794921\\
4	2.78874912090351\\
5	2.81239992667304\\
6	2.83946073093136\\
7	2.86125838078285\\
8	2.88150092515325\\
9	2.89753172268063\\
10	2.91024700589415\\
};
\addlegendentry{$d \simeq 119$ \si{\meter}, $\sigma_{\rm sh,A}= 7$ \si{\decibel}}

\end{axis}
\end{tikzpicture}%